\documentclass[conference,10pt,letterpaper]{IEEEtran}
\IEEEoverridecommandlockouts

\usepackage{cite}

\ifCLASSINFOpdf
\else
\fi
\usepackage{amsmath,amssymb,amsthm,amsfonts,amscd,amsbsy,amsxtra,amsfonts}
\usepackage{mathtools}
\DeclareMathAlphabet{\mathpzc}{OT1}{pzc}{m}{it} 
\usepackage[hyphens]{url}

\usepackage[dvipsnames]{xcolor}
\usepackage[normalem]{ulem}
\usepackage{array, makecell, cellspace}
\usepackage{enumitem}
\usepackage{svg}
\usepackage{amsthm}

\DeclareRobustCommand{\erase}{\bgroup\markoverwith{\textcolor{red}{\rule[.5ex]{2pt}{0.5pt}}}\ULon}

\DeclareMathOperator{\sinc}{Sinc}
\DeclareMathOperator{\rect}{\Pi}

\begin{document}

\bstctlcite{IEEEexample:BSTcontrol}
%

\title{A Primer on Orthogonal Delay-Doppler Division Multiplexing (ODDM)\vspace{-0.3em}}
%
%
%

\author{\IEEEauthorblockN{Hai Lin}
  \IEEEauthorblockA{Osaka Metropolitan University \\ Sakai, Osaka, 599-8531, Japan \vspace{-2.5em}}
  \thanks{
    The supplementary materials for this paper are available at:  \protect\url{https://oddm.io}
  }
  }

\maketitle


\begin{abstract}
  As a new type of multicarrier (MC) scheme built upon the recently discovered delay-Doppler domain orthogonal pulse (DDOP), orthogonal delay-Doppler division multiplexing (ODDM) aims to address the challenges of waveform design in linear time-varying channels. In this paper, we explore the design principles of ODDM and clarify the key ideas underlying the DDOP. We then derive an alternative representation of the DDOP and highlight the fundamental differences between ODDM and conventional MC schemes. Finally, we discuss and compare two implementation methods for ODDM.
\end{abstract}

\section{Introduction}

In digital communications, the role of a modulation scheme is to map digital information onto an \emph{analog waveform} that matches the characteristics of the physical channel\cite{dc4th}. 

Let $f_c$ and $\theta_c$ be the carrier frequency and phase, respectively. Typically, a \emph{real-valued} passband digital modulation waveform with symbol index $i\in \mathbb Z$ is represented as:
\begin{equation}
x_{pb}(t)=\sum_i \sqrt{2} A_i \cos(2\pi (f_c+f_i(t))t+\theta_c+\theta_i)p_i(t),\vspace*{-0.5em}
\end{equation}
where the amplitude $A_i$, the phase $\theta_i$, the frequency (possibly time-varying) $f_i(t)$ and the \emph{finite} duration pulse $p_i(t)$ can be used for modulation. The corresponding \emph{complex-valued} baseband waveform (assuming ideal carrier synchronization) is given by $x(t)=\sum_i x_i(t)$, where the $i$th symbol  
\begin{equation}\label{x_it}
x_i(t)=A_i e^{j\theta_i} e^{j2\pi f_i(t) t}p_i(t)\triangleq X[i]g_i(t)\vspace*{-0.3em}
\end{equation} 
consists of two components: an information-bearing \emph{digital symbol} (complex number) $X[i]=A_i e^{j\theta_i}$,  drawn from a signaling alphabet, and a {finite-energy, continuous-time function} $g_i(t)=e^{j2\pi f_i(t) t}p_i(t)$, referred to as \emph{transmit (modulating) pulse}.
As $x_i(t)$ can be generated by feeding $X[i]\delta(t)$ into a filter with impulse response $g_i(t)$, $g_i(t)$ is also called a \emph{transmit filter}. 
In the receiver, a correlator performing cross-correlation with a \emph{receive pulse} $\gamma_i(t)$ is usually used to extract the $i$th digital symbol as $Y[i]=\int_{-\infty}^{+\infty} y(t)\gamma_i^*(t)dt$, where $y(t)$ is the received waveform. 
Similarly, because the correlator can be implemented by passing $y(t)$ through a matched filter $\gamma_i^\ast(-t)$ and then sampling the filter output at an appropriate time\cite{dc4th,fdc}, $\gamma_i^*(-t)$ is called a \emph{receive filter}. 

Since $x(t)$ is composed of digitally modulated pulses, transmit pulses $\{g_i(t), i\in \mathbb Z\}$ play a fundamental role in waveform design and are essential to \emph{defining} the waveform. 
To avoid inter-symbol interference (ISI), the pulses are required to be mutually orthogonal. Hence, designing a digital modulation waveform is essentially equivalent to determining a set of \emph{orthogonal pulses}, also referred to as \emph{basis functions}.

\section{Waveform Design Principles}
Naturally, the first step is to determine the channel characteristics to which the orthogonal pulses are intended to match. 
For a linear time-invariant (LTI) channel, the primary characteristic is the available bandwidth $\mathbb B$. In single-carrier (SC) modulation with $f_i(t)= 0$, each transmit pulse occupies the entire bandwidth and is multiplexed in time according to a \emph{symbol interval} not less than $\frac{1}{\mathbb B}$ or equivalently a \emph{symbol rate} not greater than $\mathbb B$ to achieve the orthogonality.
Examples, including Nyquist and root Nyquist pulses whose Nyquist interval is the symbol interval, are essentially \emph{band-limited impulses}. 
The channel’s time dispersion generally disrupts the orthogonality among these pulses and introduce ISI into the correlator’s output $Y[i]$. Consequently, a digital channel equalizer is required to recover $X[i]$. 

Considering their own orthogonality and simple scaled channel output, a better choice of pulses may be the eigenfunctions of LTI channels, namely complex sinusoids $e^{j2\pi f t}, \forall f, -\infty <t < +\infty$, which obviously must be truncated for practical use. Given a truncation window function or equivalently a \emph{prototype pulse} $g(t)$, to retain orthogonality among truncated complex sinusoids, we can set $f$ to be integer multiples of a \emph{fundamental frequency} $\Delta F$, and obtain multiple pulses multiplexed in frequency according to $\Delta F$. Further considering a time interval $\Delta T$ for time-multiplexing, we can replace the symbol index $i$ with a time index $m$ and a frequency index $n$, and let $f_i(t)=n\Delta F$ to obtain a time-frequency (TF) two-dimensional (2D) multiplexing waveform

\vspace{-4mm}
\small
\begin{equation}\label{xt}
  x(t)=\sum_{m=0}^{M-1}\sum_{n=-N/2}^{N/2-1} X[m,n]\overbrace{e^{j2\pi n\Delta F (t-m\Delta T)}g(t-m\Delta T)}^{g_{m,n}(t)},\vspace{-1.5mm}
\end{equation}
\normalsize
which is popularly known as multicarrier (MC) modulation, where $N$ and $M$ are the numbers of complex sinusoids and MC symbols, respectively. Since $e^{j2\pi n\Delta F t}$ is often called \emph{subcarrier} or \emph{tone}, the transmit pulses $g_{m,n}(t)$ in (\ref{xt}) are truncated or pulse-shaped (PS) subcarriers, corresponding to \emph{rectangular} or \emph{non-rectangular} $g(t)$, respectively. Then, the channel bandwidth is matched by 

\vspace*{-3.5mm}
\begin{equation}
  B_x = (N-1)\times \Delta F+B_g\le \mathbb B \label{B_x},\vspace*{-1mm}
\end{equation}
where $B_x$ and $B_g$ are the bandwidths of $x(t)$ and $g(t)$, respectively. 
Note that the bandwidth of a finite duration signal can be reasonably defined by disregarding its negligibly small high-frequency components.

The symbol interval $\Delta T$ and the fundamental frequency (a.k.a. \emph{subcarrier spacing}) $\Delta F$ form a TF grid structure in the TF domain, where the TF resolution $(\Delta T, \Delta F)$ is meant to represent the minimum time and frequency ``distances" among the pulses - \emph{not their actual duration and bandwidth}. For such a TF domain MC (TFMC)  waveform, we have a well-known design framework based on the Weyl-Heisenberg (WH) frame theory\cite{tff}, because $g_{m,n}(t)=g(t-m\Delta T)e^{j2\pi n\Delta F (t-m\Delta T)}$ in (\ref{xt}) is exactly the WH or Gabor function widely adopted in TF signal analysis\cite{ftfa}.

A WH set $\left\{\Delta T, \Delta F, g(t)\right \}$ is said to be \emph{orthogonal} given the inner product
$\langle g_{m,n}(t), g_{\dot m, \dot n}(t)\rangle=\delta (m-\dot m)\delta (n-\dot n)$. Let another WH function $r_{m,n}(t)=r(t-m\Delta T)e^{j2\pi n\Delta F (t-m\Delta T)}$ be the corresponding receive pulse. A pair of WH sets
$\left\{\Delta T, \Delta F, g(t)\right \}$ and $\left\{\Delta T, \Delta F, r(t)\right \}$ is said to be \emph{biorthogonal} when
$\langle g_{m,n}(t), r_{\dot m, \dot n}(t)\rangle=\delta (m-\dot m)\delta (n-\dot n)$.
According to the WH frame theory, when the joint TF resolution $\Delta R=\Delta T \Delta F$ is not less than $1$, we can find (bi)orthogonal WH sets to assemble the MC waveform.
Then, the eigenfunction-based transmission for LTI channels can be realized\cite{fdc} to have $Y[n,m]\approx H(n\Delta F) X[n,m]$, where $H(n\Delta F)$ is the corresponding channel frequency response, and subsequently significantly simplify the channel equalizer. 

Let $\Pi_{\mathcal T}(t)$ stand for the rectangular pulse of length $\mathcal T$, while $T_g$ and $T_{cds}$ denote the duration of $g(t)$ and the channel delay spread, respectively. The most well-known of such MC schemes is orthogonal frequency division multiplexing (OFDM)\cite{fdc,cofdm}, where $\Delta T =T_g\ge T_{sp}+T_{cds}, g(t)=\Pi_{T_g}(t), r(t)=\Pi_{T_{sp}}(t)$, and $T_{sp}=\frac{1}{\Delta F}$ is known as \emph{symbol period}. Meanwhile,
the WH frame theory also shows that there is no (bi)orthogonal WH set when $\Delta R<1$.

On the other hand, a classic measurement of TF occupancy of $g(t)$ is its TF area (TFA) given by $A_g=\alpha T_g \alpha B_g$, where $\alpha T_g$ and $\alpha B_g$ are
the standard deviations of $g(t)$ and its Fourier transform $G(f)$, respectively. Since $g(t)$ and $G(f)$ are related via the Fourier transform, the TF domain for pulse design is an \emph{interdependent} 2D domain, and is fundamentally different from that for assembling the MC waveform, which has a TF grid structure to translate a prototype pulse \emph{already available}. 
According to the uncertainty principle, the TFA obeys a lower bound $A_g \ge \frac{1}{4\pi}$ called the Gabor limit \cite{gabor}. Therefore, there is \emph{no} 2D impulse that is ``narrow" in time and frequency simultaneously (TF ``narrow"), to be confined in a TF grid with small $\Delta T$ and $\Delta F$.

The WH frame theory and the uncertainty principle imply that an MC modulation having a \emph{fine} TF resolution may not be possible because of the lack of pulse.

\section{Waveform Design for LTV channels}
For linear time-varying (LTV) channels, similarly, the characteristics for pulse design must first be determined. Unlike LTI channels having elegant and \emph{common} complex sinusoids eigenfunctions to facilitate pulse design, the underspead LTV channels at best have a structured set of \emph{approximate} eigenfunctions, which are even \emph{channel-dependent}\cite{eigen}. It is therefore impractical to realize eigenfunction-based transmission over LTV channels.

On the other hand, in addition to the bandwidth, another primary characteristic of the channels is the available time $\mathbb T$, corresponding to the overall duration of the waveform. For LTI channels, we used to not pay attention to $\mathbb T$, because the channel is time-invariant, thus independent of $\mathbb T$. However, $\mathbb T$ is as important as the bandwidth $\mathbb B$ for \emph{time-dependent} LTV channels. In fact, during an appropriate $\mathbb T$, an LTV channel can be represented by a \emph{deterministic} delay-Doppler (DD) spread function\cite{bello}. Given a practical waveform bounded by bandwidth $\mathbb B$ and duration $\mathbb T$, the waveform received after receiving (or anti-aliasing) filtering can be sampled at an appropriate rate $B\le \mathbb B$ over a period of time $T \le \mathbb T$.
The waveform may then be considered to have experienced an equivalent sampled DD (ESDD) channel associated with a delay resolution $\frac{1}{B}$ and a Doppler resolution $\frac{1}{T}$\cite{bello}.
As a result, the waveform design for LTV channels may be replaced by that for ESDD channels, which have common characteristics of $B$ and $T$ or equivalently the DD resolution $(\frac{1}{B}, \frac{1}{T})$.

Note that \emph{physical units} of delay and Doppler are time and frequency, respectively. The DD resolution is exactly a TF resolution and therefore the waveform considering the DD resolution $(\frac{1}{B}, \frac{1}{T})$ or the so-called DD domain modulation is naturally an MC modulation. However, due to the joint TF resolution $\frac{1}{BT}\ll 1$ in practice, such DD domain MC (DDMC) modulation does not exist within the conventional TFMC waveform design framework mentioned before.

\begin{figure}
  \centering
  \includegraphics[width=8.5cm]{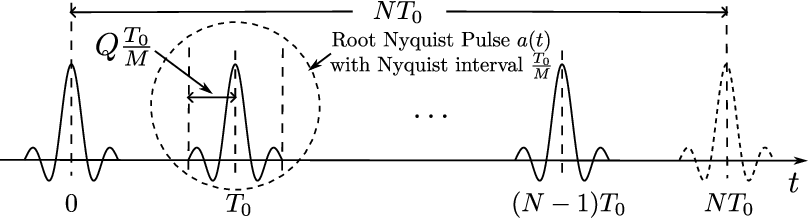}\vspace{-3mm}
  \caption{The DDOP $u(t)$.}
  \label{utfig}
  \vspace{-5mm}
\end{figure}

\section{ODDM Modulation}
As a novel MC modulation beyond the conventional TFMC waveform design framework, ODDM is based on a special DD domain orthogonal pulse (DDOP)\cite{oddmicc22,ddop,oddm,lin_mc_2023}. 
In particular, given a design parameter $T_0$, we set $\Delta T =\frac{T_0}{M}$ and $\Delta F=\frac{1}{NT_0}$ in (\ref{xt}) for ODDM with $B =\frac{M}{T_0}$ and $T=NT_0$. Then, the ODDM waveform without cyclic prefix can be represented as \cite{oddmicc22,ddop,oddm,lin_mc_2023}

\vspace{-3mm}
\small
\begin{equation}\label{xtoddm}
  \check{x}(t)=\sum_{m=0}^{M-1}\sum_{n=-N/2}^{N/2-1}X[m,n]e^{j2\pi\frac{n}{NT_0} \left(t-m\frac{T_0}{M}\right)}u\left(t-m\frac{T_0}{M}\right),\vspace*{-0.3em}
\end{equation}
\normalsize
where the DDOP $u(t)$ depicted in Fig. \ref{utfig} is a pulse-train defined as
\begin{equation}
  u(t)=\sum_{\dot n=0}^{N-1}a(t-\dot nT_0),\vspace*{-0.3em}
\end{equation}
whose elementary pulse $a(t)$ is a root Nyquist pulse parameterized by its Nyquist interval $\frac{T_0}{M}$ and duration $2Q\frac{T_0}{M}$. It has been proved in \cite{oddmicc22,oddm} that when $2Q\ll M$, although $\frac{1}{BT}=\frac{1}{MN}\ll 1$, $u(t)$ satisfies the orthogonality property of
\begin{equation}
  \mathcal A_{u,u}\left(m\frac{T_0}{M}, n\frac{1}{NT_0}\right)= \delta(m)\delta(n),\vspace*{-0.2em} \label{local_bio}
\end{equation}
for $|m|\le M-1$ and $|n| \le N-1$, where $\mathcal A_{u,u}(\cdot)$ is the ambiguity function of $u(t)$ given by the inner product
$
  \mathcal A_{u,u}(\tau,\nu) =   \langle u(t), u(t-\tau)e^{j2\pi \nu (t-\tau)} \rangle$.

\subsection{Local or Sufficient (Bi)Orthogonality}

One can see that the orthogonality in (\ref{local_bio}) is subject to a limited number of pulses.
On the other hand, the (bi)orthogonality in the conventional TFMC waveform design framework
is subject to $\left\{\Delta T, \Delta F, g(t)\right \}$ and $\left\{\Delta T, \Delta F, r(t)\right \}$, which are WH \emph{full} sets containing \emph{unlimited} number of pulses corresponding to the entire TF domain. Clearly, such \emph{global (bi)orthogonality} is not necessary for the design of a practical MC waveform, which contains a limited number of pulses and only occupies a limited region in the TF domain.

Without loss of practicality, the global (bi)orthogonality for a pair of WH full sets can be replaced by the (bi)orthogonality for a pair of WH \emph{subsets} $\left\{\Delta T, \Delta F, g(t), M,N\right\}$ and $\left\{\Delta T, \Delta F, r(t), M, N\right \}$. Since we merely need $MN$ pulses to carry $MN$ digital symbols, the WH subsets and the associated \emph{local (bi)orthogonality} are \emph{sufficient} to design the MC waveform.
Also, because the WH frame theory is applied \emph{only} to WH full sets, by introducing the local or sufficient (bi)orthogonality for WH subsets, the DDOP-based ODDM can ``bypass" the restrictions imposed by the WH frame theory, and therefore goes beyond the conventional TFMC waveform design framework to have \emph{fine} TF or DD resolution.

\subsection{Key Ideas of DDOP}
Recall that bandwidth and duration are two primary channel characteristics that determine the delay and Doppler resolutions, respectively. The DDOP considering the DD resolution is actually designed according to the bandwidth and duration of the channel.

\begin{figure}
  \centering
  \includegraphics[width=8.5cm]{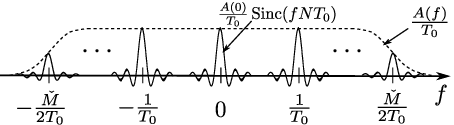}\vspace{-1mm}
  \caption{Frequency domain representation $U(f)$ of the DDOP $u(t)$.}
  \label{uffig}
  \vspace{-4mm}
\end{figure}

Let $A(f)$ be the Fourier transform of $a(t)$, the frequency domain representation of the DDOP is given by \cite{ddop}
\begin{equation}\label{ufeq}
  U(f) = \frac{e^{-j2\pi f \widetilde T}}{T_0} A(f)  \sum_{m=-\infty}^{\infty} e^{j2\pi \frac{m(N-1)}{2}}\sinc(fNT_0-mN),
\end{equation}
where $\widetilde T= \frac{1}{2}(2Q\frac{T_0}{M}+(N-1)T_0)$ and $\sinc(\cdot)$ stands for the normalized $\sinc$ function.  As depicted in Fig. \ref{uffig}, $U(f)$ consists of $(\check M+1)$ scaled $\sinc$ functions spaced by $1/T_0$ and centered from $-\frac{\check M}{2T_0}$ to $\frac{\check M}{2T_0}$ under the envelope of $A(f)$, where $\check M \ge M$ depends on the width of $A(f)$. From (\ref{ufeq}), one can see that the bandwidth of the DDOP is identical to that of the elementary pulse $a(t)$, which is known as a band-limited impulse that occupies the entire bandwidth in the SC modulation. Meanwhile, as seen in Fig. \ref{utfig}, the DDOP also has a duration close to that of the channel.
Due to the duality between time and frequency, the DDOP may also be considered as a \emph{time-limited frequency domain impulse}. These observations imply that the DDOP looks like a 2D impulse.

However, as established by the uncertainty principle, a \emph{true} 2D impulse does not exist. Does the DDOP violate the uncertainty principle? The answer is no.
Recall that the TF domain for pulse design is an interdependent 2D domain.
Due to their interdependence, a signal that is narrow in time tends to be wide in frequency, and vice versa.
A pulse that is TF ``narrow", like the non-existent 2D impulse, may also be considered as TF ``wide". Although the uncertainty principle sets a lower bound for TFA, there is \emph{no upper bound} of TFA. In fact, we can have TF ``wide" pulses using the pulse-train structure.

The key properties of the DDOP lie exactly in its pulse-train structure, particularly in the spacing among the elementary pulses.
Given the time domain spacing $T_0$, the bandwidth and duration of a pulse-train are determined by the bandwidth and the number of elementary pulses, respectively. They are independent and can be ``wide" simultaneously, as long as the pulse-train structure holds.
One can see that $U(f)$ is also a \emph{frequency domain pulse-train} windowed by $A(f)$, where the frequency domain elementary pulses $\sinc(fNT_0-mN)$ are spaced by $1/T_0$. 

Similar to (\ref{B_x}),  from (\ref{xt}), we have the duration of $x(t)$ as
\begin{align}
  T_x =  (M-1)\times \Delta T+T_g\le \mathbb T \label{T_x}.\vspace*{-0.6em}
\end{align}
In the conventional TFMC waveform design framework, $T_g$ and $B_g$ are comparable to $\Delta T$ and $\Delta F$ as $T_g \ge \Delta T$ in (\ref{T_x}) and $B_g > \Delta F$ in (\ref{B_x}), respectively. It can be seen that (\ref{B_x}) and (\ref{T_x}) are also satisfied by setting $T_g\gg \Delta T$ and $B_g\gg \Delta F$, which require the prototype pulse to be TF ``wide" and allows for staggering in both time and frequency domains.

The DDOP precisely exhibits these characteristics, featuring wide bandwidth, long duration, and the internal TF spaces among its elementary pulses. Without violating the uncertainty principle, the DDOP may be considered as a \emph{pseudo} 2D impulse, which behaves like the non-existent 2D impulse within a TF region of $(T_0, \frac{1}{T_0})$ w.r.t. the resolution of $(\frac{T_0}{M},\frac{1}{NT_0})$. Thus, the DDOP-based ODDM achieves a pseudo-2D-impulse-based transmission over ESDD channels, 
where the internal TF spacing of the DDOP results in a staggered signal structure of ODDM in both the time and frequency domains\cite{iccc24talk}.

\subsection{Alternative Representation of DDOP}
As shown in \cite{lin_mc_2023,ddop}, the DDOP can be obtained by applying a rectangular window $\rect_{NT_0}\left(t+\frac{T_0}{2}\right)$ to a periodic impulse train
$
  \ddot u(t)=\sum_{n=-\infty}^{\infty} \delta (t-nT_0),
$
and then filtering the truncated impluse train using a filter with the impulse response $a(t)$, where the order of windowing and filtering can be exchanged.
As a result, it is clear that the DDOP is \emph{time-limited and band-limited periodic impulse train}.

Because the frequency domain representation of $\ddot u(t)$
is a Fourier series, it can also be written as a frequency domain periodic impulse train having infinite number of frequency bins as
$
  \ddot U(f)=\frac{1}{T_0}\sum_{m=-\infty}^{\infty} \delta (f-\frac{m}{T_0}),
$
which indicates that $\ddot u(t)$ can be rewritten as
\begin{align}
  \ddot u(t)=\sum_{n=-\infty}^{\infty} \delta (t-nT_0)=\frac{1}{T_0}\sum_{m=-\infty}^{\infty} e^{j2\pi \frac{m}{T_0}t}.
\end{align}

By passing $\ddot u(t)$ through $a(t)$ and then applying the rectangular window, we have another representation of the DDOP given by

\vspace{-3mm}
\small
\begin{align}\label{utf}
  u(t)=\frac{1}{T_0}\left(\sum_{m=-\frac{\check M}{2}}^{\frac{\check M}{2}} A\left(\frac{m}{T_0}\right)e^{j2\pi \frac{m}{T_0}t}\right)\times \rect_{NT_0}\left(t+\frac{T_0}{2}\right),
\end{align}
\normalsize
where $(\check M+1)$ frequency bins or complex sinusoids are spaced by $1/T_0$ under the envelope of $A(f)$. It is interesting to observe that by letting
$a(t)=\sinc(\frac{Mt}{T_0})$, we have $\check M=M$ and $A(m/T_0)=A(0), \forall m$ in (\ref{utf}). Then, the summation over $m$ represents a \emph{Dirichlet kernel}, which converges to a periodic train of $\sinc(\frac{Mt}{T_0})$ with period $T_0$, as $M\gg 2Q$\cite{os3_sasp}.

Meanwhile, it should be noted that the elementary pulse $a(t)$ in the DDOP $u(t)$ can be a Nyquist pulse\cite{oddm}, such as the $\sinc(\frac{Mt}{T_0})$ pulse. As mentioned in \cite{oddm}, the purpose of using the root Nyquist pulse is to easily achieve a matched filtering at the receiver. When we use a Nyquist pulse as $a(t)$, the matched filtering may be simplified to the sampling.

\subsection{ODDM versus Other MC Schemes}

The DD domain modulation was first considered in form of orthogonal time frequency space (OTFS) modulation \cite{otfs_wcnc_2017}.
Due to the absence of 2D impulse\cite{hadaniyt}, the OTFS transforms digital symbols from the DD domain to the TF domain via the so-called inverse symplectic finite Fourier transform. The transformed digital symbols are then modulated using the conventional OFDM or TFMC waveforms. Therefore, OTFS can be considered as a precoded OFDM\cite{zemenpimrc2018}, which is within the conventional TFMC waveform design framework.

For comparison purposes, let the TF domain grid of OTFS obey $\{\hat nT_0,\hat m\frac{1}{T_0}\}$, then the corresponding DD domain grid is considered to be $\{m\frac{T_0}{M}, n\frac{1}{NT_0}\}$.
The comparison of ODDM, OFDM, and OTFS is shown in Table I. One can see that ODDM is fundamentally different from OFDM and OTFS, in terms of physical concepts, transmission strategy, waveform design framework, and the resultant orthogonality.

\setcellgapes{0.5ex}\makegapedcells
\begin{table*}
  \centering
  \caption{Comparison of OFDM, OTFS, and ODDM}
  \begingroup
  \setlength{\tabcolsep}{0.4em}
  \begin{tabular}{|l|c|l|l|}
    \hline
    \thead{}                            & \thead{OFDM}                                            & \thead{OTFS} & \thead{ODDM} \\
    \hline
    \makecell{Physical                                                                                                          \\ concepts} & TF domain & \makecell[l]{ DD and TF domains are different domains. \\ DD domain modulation is challenging, \\ due to the absence of 2D impulse\cite{hadaniyt}.} & \makecell[l]{ A DD domain in practice is a TF domain with \\ fine TF resolution. A practical DD domain \\ modulation is an MC modulation \cite{oddmicc22,oddm}.} \\
    \hline
    \makecell{Transmission                                                                                                      \\ strategy} & \makecell{Eigenfunction-based transmission                                                                                                                         \\  for LTI channels} & \makecell{Precoded OFDM/TFMC\cite{zemenpimrc2018}}      & \makecell{Pseudo-2D-impulse-based \\ transmission for LTV/ESDD channels \cite{lin_mc_2023} } \\
    \hline
    \makecell{                                                                                       Waveform
    design                                                                                                                      \\ framework} & \makecell{WH \emph{full set} bounded by \\ the WH frame theory\cite{tff}} & \makecell{WH \emph{full set} bounded by \\the WH frame theory\cite{tff}} & \makecell{WH \emph{subset} beyond the restrictions \\ imposed by the WH frame theory\cite{ddop}}  \\ 
    \hline
    \makecell{Orthogonality}             & \makecell{Global (bi)orthogonality w.r.t.                                              \\ coarse TF grids $\{m\Delta T, n\Delta F\}$, \\ $\Delta T \ge \frac{1}{\Delta F}$, for \emph{unbounded} $m$ and $n$\cite{tff}} & \makecell{Global (bi)orthogonality w.r.t. \\ coarse TF grids $\{\hat nT_0, \hat m\frac{1}{T_0}\}$ \\ for \emph{unbounded} $\hat n$ and $\hat m$ \cite{tff}} & \makecell{Local or sufficient  (bi)orthogonality w.r.t. \\ fine TF grids $\{m\frac{T_0}{M}, n\frac{1}{NT_0}\}$ \\ for \emph{bounded} $m$ and $n$\cite{ddop}} \\
    \hline
  \end{tabular}
  \endgroup
\end{table*}

\section{Implementation Methods for ODDM}
Being an MC scheme with non-rectangular prototype pulse, ODDM is a pulse-train-shaped (PTS) OFDM and therefore can be ordinarily implemented as a PS-OFDM. In addition, due to the unique pulse-train structure of the prototype pulse $u(t)$, it can also be \emph{approximately} implemented as a \emph{wideband} filtered OFDM\cite{oddm}.
\subsection{PS-OFDM Based Implementation}
For ODDM signal in (\ref{xtoddm}), we have the $m$th symbol

\vspace{-3mm}
\small
\begin{equation}\label{check_xmt}
  \check{x}_m(t)=\sum_{n=-N/2}^{N/2-1}X[m,n]e^{j2\pi\frac{n}{NT_0} \left(t-m\frac{T_0}{M}\right)}u\left(t-m\frac{T_0}{M}\right), \vspace*{-0.5em}
\end{equation}\normalsize
which is
$
  \dot {x}_m(t)=\sum_{n=-N/2}^{N/2-1}X[m,n]e^{j2\pi\frac{n}{NT_0} \left(t-m\frac{T_0}{M}\right)}
$ pulse-shaped (truncated) by $u(t)=\sum_{\dot n=0}^{N-1}a(t-\dot nT_0)$. 
Clearly, $\dot x_m(t)$ is an \emph{infinite periodic} signal with period $NT_0$ and is strictly band-limited to $[-\frac{1}{2T_0}, \frac{1}{2T_0}]$. Therefore, $\dot x_m(t)$ can be sampled at a rate of $1/T_0$ to have $\dot {\mathbf x}_m= [\dot x_m[0], \cdots, \dot x_m[N-1]]^T$, which represents $N$ samples sampled every $T_0$ within one period of $NT_0$.

Since $\dot {\mathbf x}_m$ is exactly the inverse discrete Fourier transform (IDFT) of $[X[m,0],\cdots,X[m,\frac{N}{2}-1],X[m,-\frac{N}{2}],\cdots,$ $X[m,-1]]^T$, we can pass \emph{cyclic extension} (CE) of $\dot {\mathbf x}_m$ denoted by 
$
  \dot {\mathbf x}_m^{CE}=[\cdots, \dot {\mathbf x}_m^T, \dot {\mathbf x}_m^T, \dot {\mathbf x}_m^T \cdots]^T
$ through ideal low-pass filter (LPF) with cut-off frequency $\frac{1}{2T_0}$ (a.k.a. interpolation filter) to obtain $\dot x_m(t)$ \cite{oddmicc22}. After that, the $u(t)$-based pulse-shaping is applied to $\dot x_m(t)$ to generate $\check x_m(t)$ in (\ref{check_xmt}) as
\begin{equation}\label{check_idft_xmt}
  \check x_m(t)=\overbrace{\left(\dot {\mathbf x}_m^{CE}*\sinc\left(\frac{t}{T_0}\right)\right)}^{\dot x_m(t)} \times u\left(t-m\frac{T_0}{M}\right)
\end{equation}
where $*$ stands for convolution. It should be noted that, since the bandwidth of $\check x_m(t)$ is greater than $\frac{1}{T_0}$, neither $\dot {\mathbf x}_m$ nor the truncated $\dot {\mathbf x}_m^{CE}$ can represent $\check x_m(t)$, due to \emph{the violation of the sampling theorem}. The above steps follow the standard IDFT-based implementation of OFDM and PS-OFDM\cite{lin_mc_2023}.

\subsection{Filtered OFDM Based Approximate Implementation}

Let $\check{x}_m^{\dot n}(t)$ denote $\check{x}_m(t)$ within the duration of $\dot n$th elementary pulse in $u\left(t-m\frac{T_0}{M}\right)$. From (\ref{check_xmt}), we observe
\begin{align}
  \check {x}_m^{\dot n}(t) & =\sum_{n=-N/2}^{N/2-1}X[m,n]e^{j2\pi\frac{n}{NT_0} (t-m\frac{T_0}{M})} a\left(t-\dot n T_0-m\frac{T_0}{M}\right), \nonumber \\
                           & \approx \sum_{n=-N/2}^{N/2-1}X[m,n]e^{j2\pi\frac{n\dot n}{N}} a\left(t-\dot n T_0-m\frac{T_0}{M}\right),  \nonumber
\end{align}
for $\dot n T_0+(m-Q)\frac{T_0}{M}\le t \le \dot n T_0+(m+Q)\frac{T_0}{M}$ when $2Q\ll M$\cite{oddm}. Then, we can obtain

\vspace{-3mm}
\small
\begin{equation}
  \check {x}_m(t) \approx \sum_{\dot n=0}^{N-1}\sum_{n=-N/2}^{N/2-1}X[m,n]e^{j2\pi\frac{n\dot n}{N}} a\left(t-\dot n T_0-m\frac{T_0}{M}\right),
\end{equation}
\normalsize
where $\sum_{n=-N/2}^{N/2-1}X[m,n]e^{j2\pi\frac{n\dot n}{N}}, 0\le \dot n \le N-1$ exactly form $\dot {\mathbf x}_m$.  
As a result, the $m$th ODDM symbol can be approximately generated by filtering  $\dot {\mathbf x}_m$ with $a(t)$ \cite{oddm}, expressed as 
\begin{equation}\label{filter_check_xmt}
  \check{x}_m(t)\approx \dot {\mathbf x}_m*a(t).
\end{equation}
Recall that the fundamental frequency of the ODDM is $\Delta F=\frac{1}{NT_0}$. For a conventional filtered OFDM \cite{fofdm,fofdmspawc} with the same fundamental frequency, the cut-off frequency of the LPF or interpolation filter is around $\frac{1}{2T_0}$. Considering $a(t)$ has a cut-off frequency greater than $\frac{M}{2T_0}$, the approximate ODDM in (\ref{filter_check_xmt}) is a wideband filtered OFDM.

\section{Waveform-level Simulations of ODDM}
The waveform-level simulation of a waveform $x(t)$ having bandwidth $B_x$ is performed using discrete-time samples $x[k]\triangleq x\left(\frac{k}{W}\right), k\in \mathbb Z$, with the sampling rate $\mathbb W \gg B_x$. Taking into account the less stringent definition of $B_x$ and the frequency dispersion of the channel, we often choose $\mathbb W$ to be $8$ to $16$ times $B_x$. Meanwhile, since the noise also needs to be represented by  $\frac{1}{\mathbb W}$-spaced samples, its power is $N_0\mathbb W$, where $N_0$ is its one-sided power spectral density. 
Because the noise power depends on $\mathbb W$, we prefer $\frac{E_b}{N_0}$ to the signal-to-noise ratio in waveform-level simulations, where the bit energy $E_b$ is determined by the energy of $g_i(t)$ and the average power of the signaling alphabet in (\ref{x_it}).
This simulation setup is well justified by considering a \emph{band-pass filter} or a \emph{baseband anti-aliasing LPF} in the link, which rejects signal components outside of $[-\frac{\mathbb W}{2}, \frac{\mathbb W}{2}]$ to satisfy the sampling theorem.

\begin{figure}
  \centering
  \includegraphics[width=8.5cm]{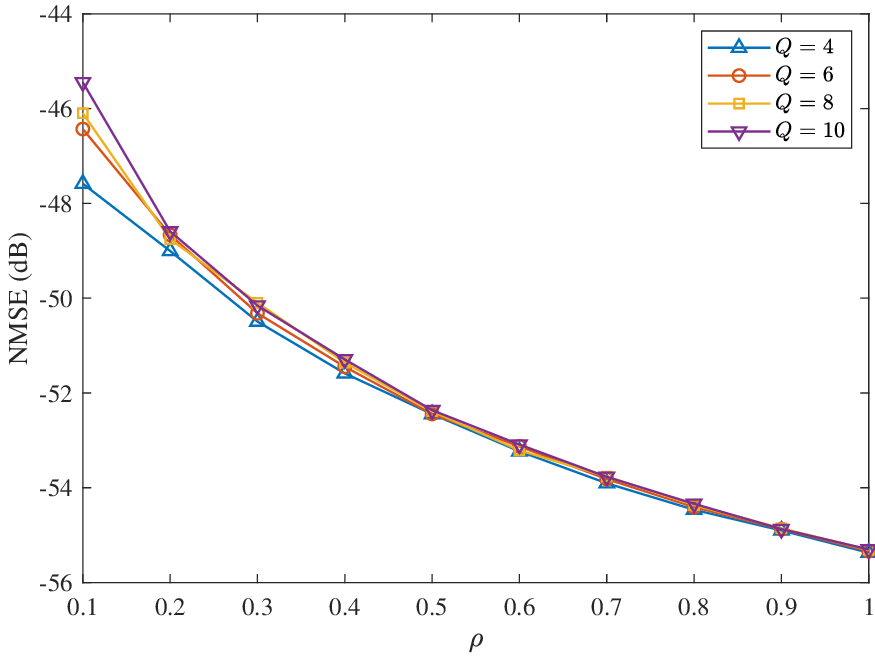}\vspace*{-3mm}
  \caption{NMSE of approximate ODDM waveform, $M=512$, $N=32$.} \vspace*{-6mm}
  \label{xt_appro_nmse}
\end{figure}

Due to space limitations, we present only the simulation results of the normalized mean square error (NMSE) between the approximate and exact ODDM waveforms in (\ref{filter_check_xmt}) and (\ref{check_idft_xmt}), respectively. In these simulations, $a(t)$ is a root-raised cosine pulse with roll-off factor $\rho$, and $\mathbb W$ is set to $8\frac{M}{T_0}$.
As shown in Fig. \ref{xt_appro_nmse}, the NMSE decreases as $Q$ decreases, while it decreases as $\rho$ increases. Since the NMSE values in all cases are below $-45$dB, it can be concluded that, \emph{with appropriate parameter settings}, the wideband filtered OFDM based implementation can effectively generate the ODDM waveform.


\vspace*{-1mm}

\section{Conclusion}
The design principles of ODDM were investigated, and the key ideas of DDOP were clarified. An alternative representation of the DDOP was derived, and fundamental differences between ODDM and conventional MC schemes were highlighted. Two implementation methods of ODDM were presented, and their effectiveness was validated through waveform-level simulations. As a novel waveform, research on ODDM remains in its early stages, and further studies are needed to fully explore its potential and address challenges related to practical deployment.

\vspace{-1.4mm}




%





\ifCLASSOPTIONcaptionsoff
  \newpage
\fi



%
\bibliographystyle{IEEEtran}
\bibliography{oddm}





%






\end{document}